\documentclass[onecolumn,aps,pra,reprint,superscriptaddress,longbibliography]{revtex4-1}

\usepackage{amsmath}
\usepackage{graphicx}
\usepackage[lofdepth,lotdepth, caption=false]{subfig}
\usepackage{verbatim}
\usepackage{color}
\usepackage{dcolumn}
\usepackage{bm}
\usepackage{miller}
\usepackage{color}

\usepackage{xr-hyper}
\usepackage{hyperref}

\usepackage[english]{babel}

\newcommand{\beginsupplement}{%
        \setcounter{table}{0}
        \renewcommand{\thetable}{S\arabic{table}}%
        \setcounter{figure}{0}
        \renewcommand{\thefigure}{S\arabic{figure}}%
     }

\begin{document}

\title{Role of stacking defects on the magnetic behavior of CrCl$_3$}

\author{John A.~Schneeloch}
\affiliation{Department of Physics, University of Virginia, Charlottesville,
Virginia 22904, USA}

\author{Adam A.\ Aczel}
\affiliation{Neutron Scattering Division, Oak Ridge National Laboratory, Oak Ridge, Tennessee 37831, USA}

\author{Feng Ye}
\affiliation{Neutron Scattering Division, Oak Ridge National Laboratory, Oak Ridge, Tennessee 37831, USA}

\author{Despina Louca}
\thanks{Corresponding author}
\email{louca@virginia.edu}
\affiliation{Department of Physics, University of Virginia, Charlottesville,
Virginia 22904, USA}

\begin{abstract}    
In the study of van der Waals-layered magnetic materials, the properties of CrCl$_3$ continue to attract attention. 
This compound is reported to undergo antiferromagnetic (AFM) ordering below $\sim$14 K, with a ferromagneticlike region proposed to exist between 14 and 17 K. 
Ideally, the crystal structure is rhombohedral (R) below $\sim$235 K, separated from a higher-temperature monoclinic (M) phase by a layer-sliding structural phase transition. 
However, the structural transition is often inhibited even in bulk single crystals, allowing M-type layer stacking, reported to have a tenfold greater interlayer magnetic coupling than R-type stacking, to be present at low temperature. 
To clarify the effect of stacking defects on CrCl$_3$, we report magnetization measurements on samples of varying crystalline quality. 
At low applied magnetic field, some crystals predominantly show the $T_N=14$ K peak, but other crystals show hysteretic behavior and a magnetization enhancement  at a slightly higher temperature ($14 < T \lesssim 17$ K.) 
Samples with anomalous behavior exhibit a transition around $\sim$2 T in isothermal magnetization-field data, providing evidence that M-type stacking defects are the source of these anomalies. 
Ground powder samples are especially likely to show strongly anomalous behavior. 
We suggest that the anomalous behavior arises from few-layer magnetic domains that form just above $T_N$ in an environment of mixed interlayer magnetic coupling strength. 
We argue that the influence of M-type stacking boundaries on sublattice magnetization is already observable in reported neutron scattering data, and may be responsible for a certain feature in reported specific heat data.
\end{abstract}

\maketitle

\section{Introduction}
In recent years, the field of van der Waals layered magnetic materials has received considerable attention. 
An intriguing property of many of these materials is the ability to modify the magnetic behavior by changing the layer stacking. 
For example, in CrI$_3$, the surprising finding of antiferromagnetic (AFM) order in bilayer samples \cite{huangLayerdependentFerromagnetismVan2017}, in contrast to the ferromagnetic (FM) order reported in the bulk \cite{mcguireCouplingCrystalStructure2015}, was soon attributed to a different kind of layer stacking corresponding to a monoclinic (M) rather than rhombohedral (R) crystal structure \cite{huangEmergentPhenomenaProximity2020}. M-type stacking defects also affect bulk properties, inducing a magnetization hysteresis \cite{schneelochAntiferromagneticferromagneticHomostructuresDirac2024}, for example.

\begin{figure}[h]
\begin{center}
\includegraphics[width=8.6cm]
{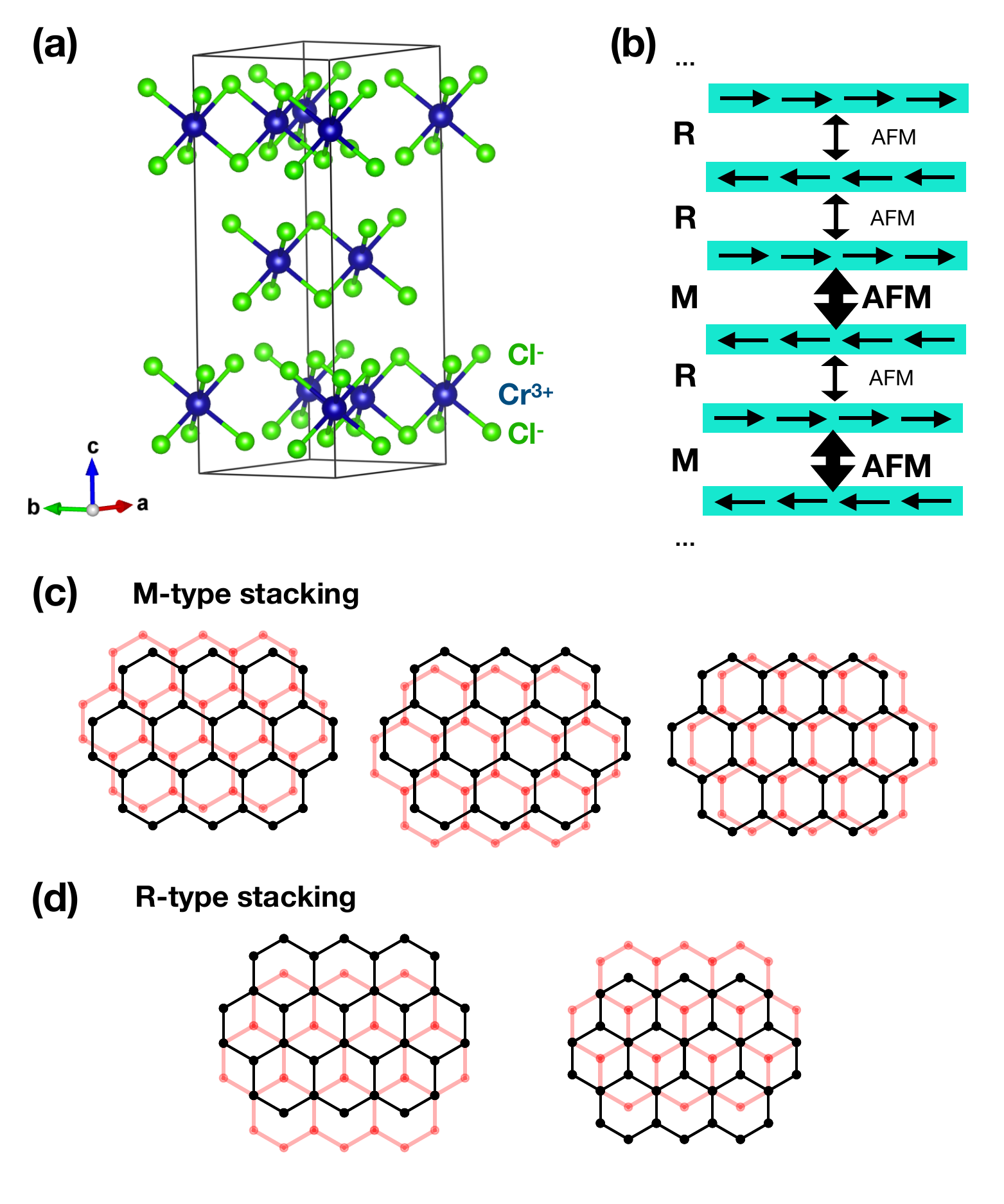}
\end{center}
\caption{(a) The crystal structure of R-phase CrCl$_3$, plotted in \textsc{VESTA} \cite{mommaVESTAThreedimensionalVisualization2011}. (b) A schematic representation of the reported tenfold greater interlayer AFM coupling between layers that are M-type rather than R-type stacked. (c,d) A depiction of the different stacking types (three for M-type (c), two for R-type (d)) as viewed perpendicular to the layers. Only the Cr$^{3+}$ ions are shown, which form honeycomb lattices.}
\label{fig:1}
\end{figure}

CrCl$_3$ is similar to CrI$_3$, differing mainly in having AFM order (below 14 K \cite{mcguireMagneticBehaviorSpinlattice2017}) rather than FM order. 
As for CrI$_3$, CrCl$_3$ ideally has a structural transition from a high-temperature phase with M-type stacking (space group $C2/m$) to a lower temperature phase with R-type stacking $(R\bar{3}$), as depicted in Fig.\ \ref{fig:1}(c,d). Ideally, the $M$$\rightarrow$$R$ structural transition occurs at 235 K on cooling and 250 K on warming, though, even in single crystals, the transition is often broadened or inhibited \cite{mcguireMagneticBehaviorSpinlattice2017}. 
The layers of CrCl$_3$ consist of a honeycomb lattice of Cr$^{3+}$ ions with $S=3/2$ spins, sandwiched by outer layers of Cl$^{-}$ ions (Fig.\ \ref{fig:1}(a).) 

The AFM order of CrCl$_3$, typically characterized in its R-phase, consists of in-plane FM alignment and inter-plane AFM alignment, with the spins laying in-plane. 
There is a weak intralayer magnetic anisotropy, with a spin-flop transition occurring around $\mu_0 H=0.01$ to 0.02 T for in-plane magnetic field \cite{mcguireMagneticBehaviorSpinlattice2017,kuhlowMagneticOrderingCrCl31982}. 
A small field of $\mu_0 H \sim 0.3$ T is sufficient to suppress AFM order \cite{mcguireMagneticBehaviorSpinlattice2017}, and, aside from shape effects, the response of CrCl$_3$ to field is very nearly isotropic \cite{narathSpinWaveAnalysisSublattice1965}. 
CrCl$_3$ has the weakest interlayer magnetic coupling of the chromium trihalides \cite{narathZeroField53MathrmCr1965}, with a magnitude hundreds of times weaker than that of the nearest-neighbor (intralayer) interaction \cite{narathSpinWaveAnalysisSublattice1965}. 

From studies on few-layer samples, which tend to exhibit M-type stacking \cite{hanAtomicallyUnveilingAtlas2023}, the interlayer magnetic coupling has been estimated to be about ten times greater for M-type than R-type stacking \cite{serriEnhancementMagneticCoupling2020, kleinEnhancementInterlayerExchange2019}, as illustrated in Fig.\ \ref{fig:1}(b). Presumably, the presence of M-type stacking would affect magnetic properties. Nuclear magnetic resonance (NMR) can measure the sublattice magnetization of CrCl$_3$ (essentially, the average magnetization within a layer); grinding samples (which would induce stacking defects) was observed to substitute the ``normal'' observed resonance with an anomalous resonance at a slightly higher frequency, indicating an increased sublattice magnetization at nonzero temperatures \cite{narathLowTemperatureSublatticeMagnetization1963}. This effect was unexplained at the time, but is consistent with the expectation that the stronger interlayer coupling of M-type stacking would make the ordering more robust. 

A close look at the literature reveals some inconsistencies in the reported behavior of CrCl$_3$. 
Specific heat data collected on pressed pellets show a broad hump near 17 K \cite{kostryukovaSpecificHeatAnhydrous1972, hansenHeatCapacitiesCrF31958} that was presumed to indicate the AFM transition, but later single crystal data \cite{mcguireMagneticBehaviorSpinlattice2017,bastienSpinglassStateReversed2019} showed a sharp peak near 14 K. 
There appears to be an inverse correlation in the size of the 14 K peak and 17 K hump; see Supplemental Materials for a comparison of specific heat data from the literature \cite{supplement}. A pressed pellet would presumably have an inhibited $M$$\rightarrow$$R$ transition and thus a substantial amount of M-type stacking at low temperature (as in CrI$_3$ \cite{schneelochAntiferromagneticferromagneticHomostructuresDirac2024}), possibly explaining the prominence of the 17 K hump in earlier studies that measured a pressed pellet. Regardless, the presence of two transitions, and especially the apparent opposite temperature shifts under applied field, has been interpreted as evidence for a ``ferromagneticlike'' phase around 14 to 17 K \cite{mcguireMagneticBehaviorSpinlattice2017,kuhlowMagneticOrderingCrCl31982}, in which intralayer FM correlations are present before 3-dimensional AFM order sets in below 14 K. 
Also within 14 to 17 K, an anomalous enhancement in the magnetic susceptibility has been observed under low applied magnetic field ($\mu_0 H \lesssim 5$ mT), with transitions speculated to be present near 16 K \cite{liuAnisotropicMagnetocaloricEffect2020,bykovetzCriticalRegionPhase2019} and 17 K \cite{bykovetzCriticalRegionPhase2019}. Ac susceptibility data suggest that the $\sim$16 K transition, unlike the 14 K transition, shows dissipative behavior  \cite{liuAnisotropicMagnetocaloricEffect2020}.

To clarify the effect of stacking defects on CrCl$_3$, we report magnetic susceptibility data on CrCl$_3$ powder and single crystal samples, focusing on their behavior at low magnetic field. 
While some crystals predominantly show the transition at 14 K, with no hysteresis or field-dependence at low field, other samples show an anomalous susceptibility enhancement around 14 to 17 K, larger on cooling than warming, larger with in-plane than out-of-plane field, and especially prominent at very low field. 
Samples with the anomalous behavior also show a transition around $\sim$2 T in isothermal magnetization-field measurements, consistent with the tenfold increase in interlayer magnetic coupling reported for M-type stacking. 
The anomalous behavior is especially strong in ground powder samples. 
A hysteresis in the isothermal magnetization-field loop opens up for temperatures within 14 to 17 K. We provide a theory for this behavior in terms of few-layer FM domains that form just above 14 K due to the disordered interlayer magnetic coupling. We discuss the effect of stacking defects on reported data including neutron scattering measurements of the sublattice magnetization, and specific heat. 

\section{Experimental Details}
CrCl$_3$ powder was purchased from Thermo Fisher Scientific (``Chromium(III) chloride, anhydrous, 99.9\% (metals basis)''). For ``unground'' powder samples, the powder was taken straight from the bottle, and the ``ground'' powder samples were ground for about a minute in a mortar and pestel. 

Crystals were grown by chemical vapor transport. CrCl$_3$ powder was sealed in an evacuated ampoule, then heated at 700 $^{\circ}$C for about a week in a single-zone tube furnace, using its natural temperature gradient to facilitate crystal growth. 

Magnetization measurements were performed in a Quantum Design Physical Property Measurement System equipped with a Vibrating Sample Magnetometer. Data taken on heating/cooling were continuously collected at a rate of 0.3 K/min unless otherwise specified. The stated values of the applied magnetic field are nominal values.

To measure the sublattice magnetization as a function of temperature, accessible via the intensity of the $(00\frac{3}{2})$ magnetic Bragg peak (Fig.\ \ref{fig:SublatticeMagnetization}), elastic neutron scattering on a single crystal was performed on the instrument VERITAS at the High Flux Isotope Reactor at Oak Ridge National Laboratory (ORNL). The instrument was operated in its standard three-axis configuration with a collimation of 40$^{\prime}$-40$^{\prime}$-40$^{\prime}$-80$^{\prime}$.
A phase diagram was constructed from additional single-crystal measurements on CORELLI \cite{yeImplementationCrossCorrelation2018} at the Spallation Neutron Source at ORNL. 
Reciprocal space locations mentioned are in reference to the R-phase unit cell of CrCl$_3$ \cite{mcguireMagneticBehaviorSpinlattice2017}.

\section{Results}
\subsection{Magnetization}

\begin{figure}[t]
\begin{center}
\includegraphics[width=8.6cm]
{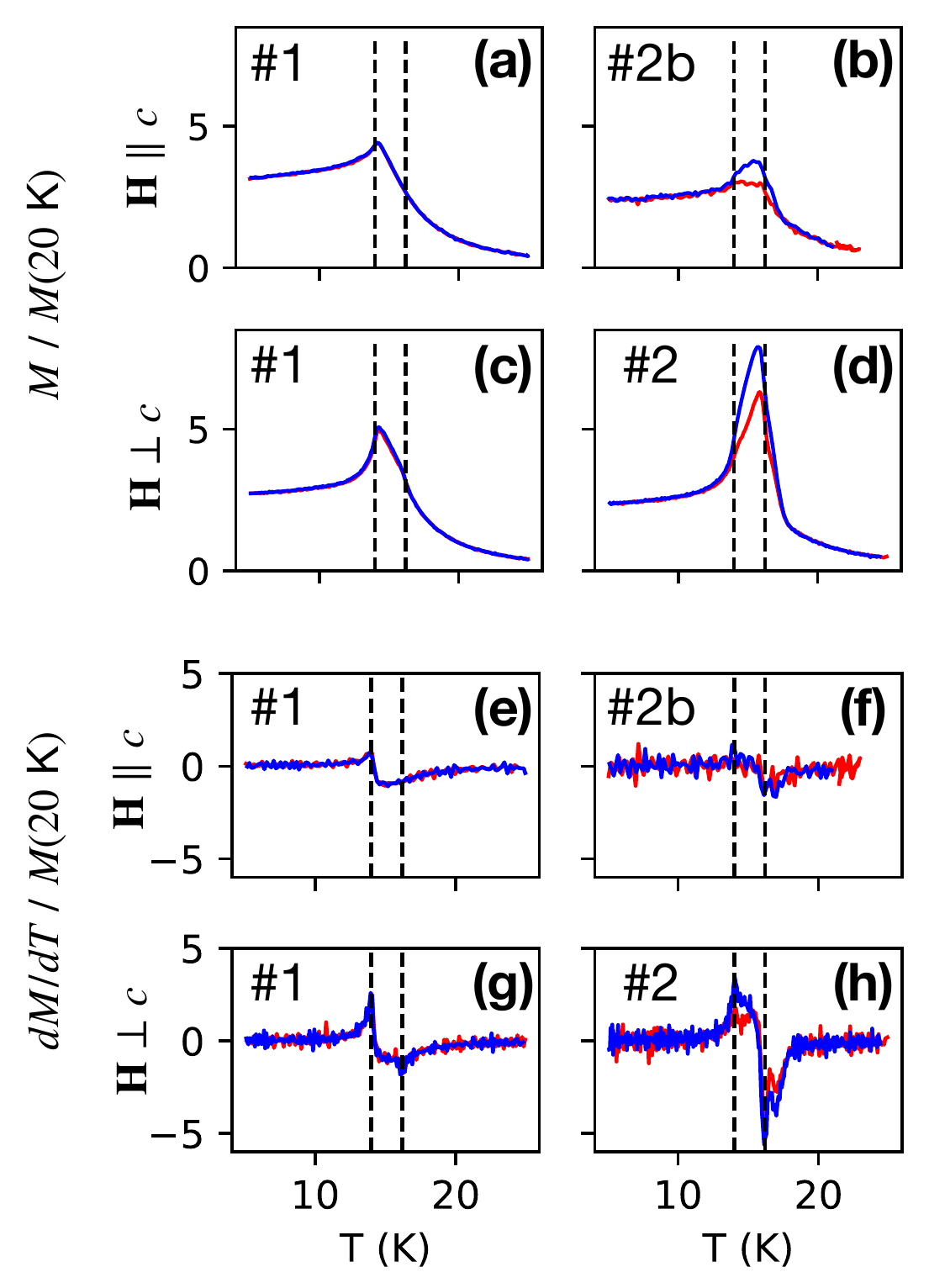}
\end{center}
\caption{(a-d) Magnetization vs.\ temperature data taken on crystal \#1 (a,c), \#2b (b), and \#2 (d), with \#2b being a piece cut from \#2. Magnetic field (nominally 0.25 mT) was applied out-of-plane (a,b) or in-plane (c,d). 
The magnetization is normalized to its intensity at 20 K. 
Red and blue lines represent ZFC heating and FC cooling, respectively. (e-h) The derivative $dM/dT$ calculated from the data in (a-d). The dashed lines are at 14.0 and 16.2 K.}
\label{fig:SuscCrystals}
\end{figure}

To show that different CrCl$_3$ crystals can have different magnetic behavior at low magnetic field, we compare magnetization data taken on two representative crystals, labeled \#1 and \#2. (Crystal \#2b is a piece cut from \#2.) Figure \ref{fig:SuscCrystals}(a-d) shows magnetization vs.\ temperature at a nominal applied field of $\mu_0 H = 0.25$ mT. Zero-field-cooled (ZFC) data were taken first, in which the crystal was cooled to 3 K before applying field and warming to 25 K, followed by field-cooling (FC) back to 3 K. For crystal \#1 in Fig.\ \ref{fig:SuscCrystals}(a,c), ZFC and FC curves overlap. 
For crystal \#1, there is a sharp kink near 14.1 K for both in-plane and out-of-plane magnetic field, though the in-plane data show a subtle shoulder near 16.0 K. These features are clearly visible in the derivative $dM/dT$, with peaks at 14.0 K and, for in-plane field, a dip at 16.2 K.

Crystal \#2, on the other hand, shows a broader peak around 14 to 17 K. There is a hysteresis with smaller magnetization on initial warming than subsequent cooling. 
There is still a kink near 14 K with a peak in $dM/dT$ (Fig.\ \ref{fig:SuscCrystals}(f,h)), but the 16 K $dM/dT$ peak is much stronger, and present for both in-plane and out-of-plane field. 
Crystal \#2 also has a broad $dM/dT$ peak near 17 K. The 16 K \cite{liuAnisotropicMagnetocaloricEffect2020, bykovetzCriticalRegionPhase2019} and 17 K \cite{bykovetzCriticalRegionPhase2019} features were noted previously, though the possibility of these anomalies arising from stacking defects was not addressed. 
Given the sample dependence of the anomalous behavior, and the presence of M-type stacking defects being the most pervasive structural variation between samples, M-type stacking defects are the most likely cause of the anomalies.

\begin{figure}[t]
\begin{center}
\includegraphics[width=8.6cm]
{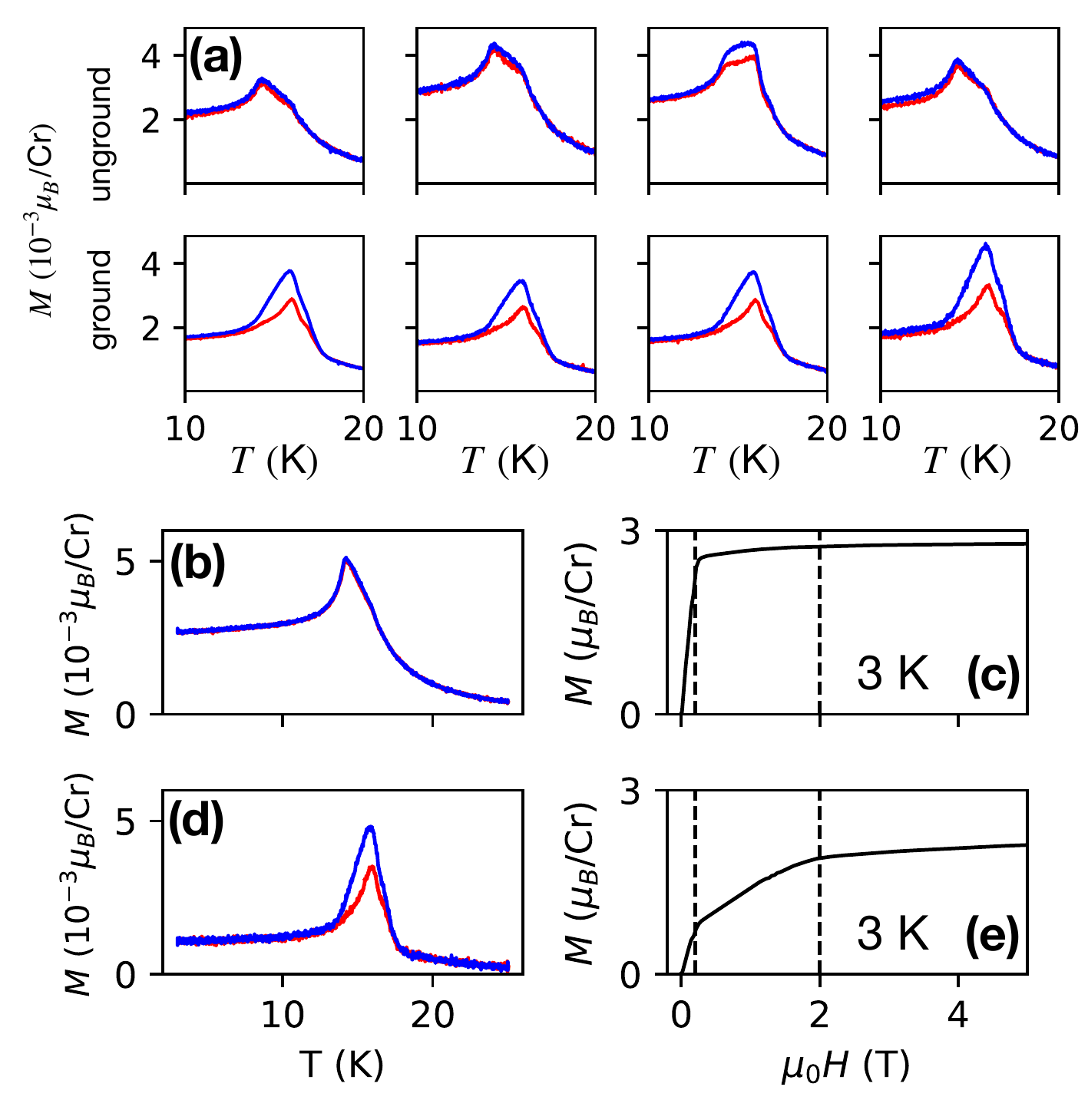}
\end{center}
\caption{(a) Magnetization vs.\ temperature of eight powder CrCl$_3$ samples taken at a nominal field of $\mu_0 H=0.25$ mT. The upper and lower rows show data from unground and ground powder samples, respectively. 
(b-e) Magnetization data taken on crystals without (b,c) and with (d,e) anomalous behavior, with field applied in-plane. (b) and (d) show magnetization vs.\ temperature at a nominal field of $\mu_0 H=0.25$ mT. (c) and (e) show magnetization vs.\ applied field at 3 K. Vertical lines show 0.2 and 2 T. 
Red and blue curves in (a,b,d) indicate ZFC (warming) and FC (cooling), respectively.}
\label{fig:grinding}
\end{figure}

Further evidence for the role of M-type stacking defects in the anomalous behavior is in their prominence in ground powder samples, and, especially, the presence of a transition near $\sim$2 T in isothermal magnetization-field measurements in samples with anomalies. 
In Fig.\ \ref{fig:grinding}(a), we show low-field magnetization data on powder samples, with or without grinding in a mortar and pestel for about a minute. The top row shows data obtained from CrCl$_3$ powder taken straight from the bottle, which happen to have minimal magnetic hysteresis. In contrast, for the ground powder, a wide magnetic hysteresis with a peak at 16 K is present in all four samples. 
These results are consistent with the assumption that grinding creates defects that inhibit the $M$$\rightarrow$$R$ layer-sliding structural phase transition, preserving M-type stacking to low temperature where they are responsible for the anomalous behavior. 

If field is applied in-plane, the AFM order is expected to be extinguished near $\mu_0 H=0.2$ T for a purely R-phase crystal, but near $\sim$2 T for a crystal with M-type stacking \cite{serriEnhancementMagneticCoupling2020}. 
In Fig.\ \ref{fig:grinding}(b,c) and (d,e), data from two crystals are compared. In Fig.\ \ref{fig:grinding}(b) and (d), low-field (0.25 mT) magnetization vs.\ temperature data for each crystal are shown, and their magnetization vs.\ field data at 3 K are shown in (c) and (e), respectively. In all four subplots, field was applied in-plane. 
For the crystal without a magnetic anomaly near 14 to 17 K (Fig.\ \ref{fig:grinding}(b)), only the 0.2 T transition is clearly present in Fig.\ \ref{fig:grinding}(c). For the crystal with a prominent hysteresis (Fig.\ \ref{fig:grinding}(d)), though, both a 0.2 T and 2 T transition are seen in Fig.\ \ref{fig:grinding}(e). 
Additional examples are shown in the Supplemental Materials \cite{supplement}, establishing a trend that samples that show the magnetization anomaly also have a prominent 2 T transition, consistent with the assumption that the anomalies are due to M-type stacking defects.

\begin{figure}[h!]
\begin{center}
\includegraphics[width=8.6cm]
{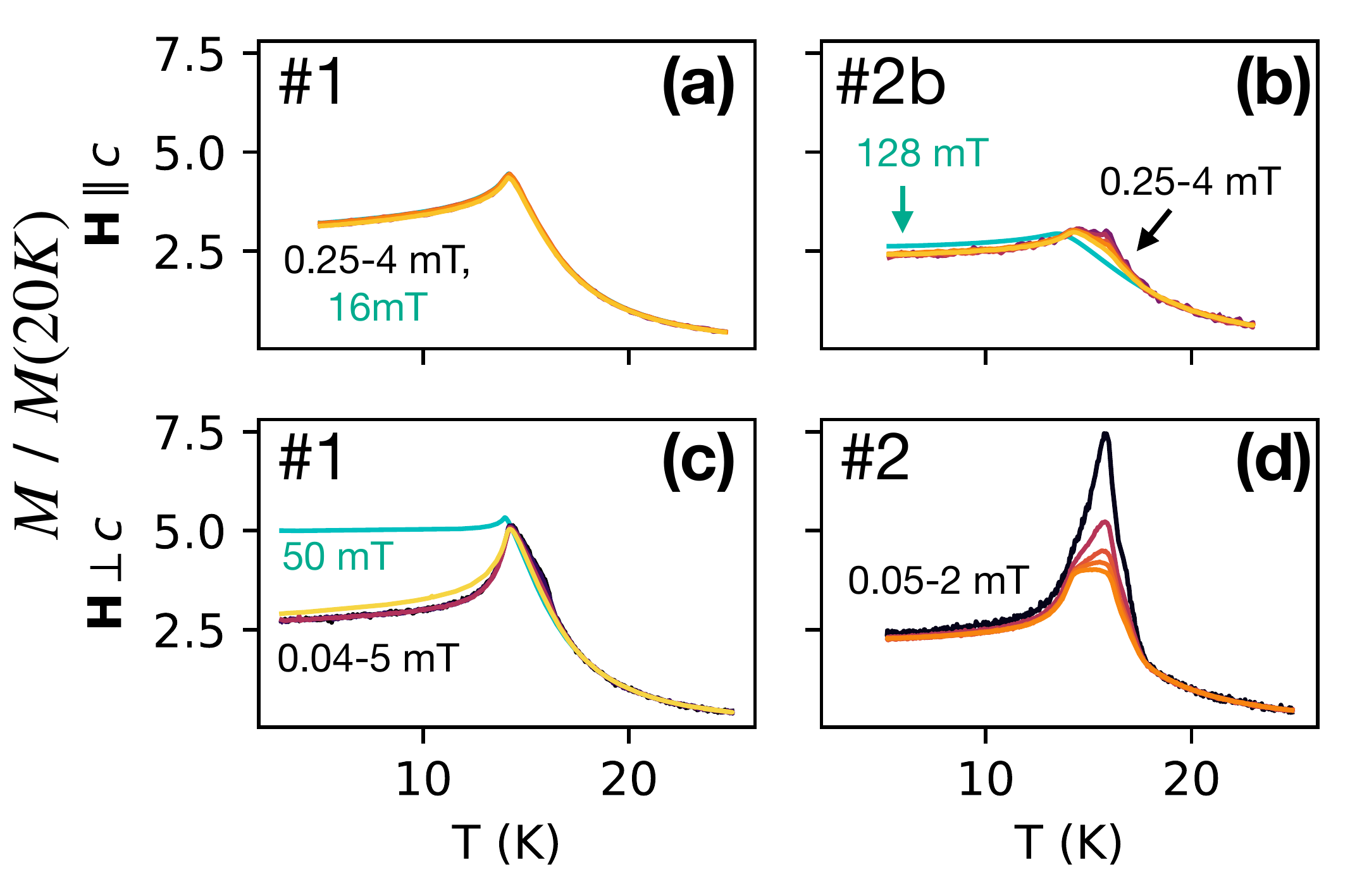}
\end{center}
\caption{Magnetization vs.\ temperature at many applied magnetic fields. 
Fields from $\mu_0 H=0.04$ to 5 mT are shown on a color scale from black to yellow. The plotted range of these fields is $\mu_0 H=0.25$ to 4 mT for (a), 0.25 to 4 mT for (b), 0.04 to 5 mT for (c), and 0.05 to 2.0 mT for (d). 
Additionally, the cyan curves show data taken at several higher magnetic fields, specifically, 16 mT for (a) (overlapped by the other data), 128 mT for (b), and 50 mT for (c). All data are normalized to their value at 20 K. For clarity, only ZFC data are shown; FC data are shown in the Supplemental Materials \cite{supplement}.}
\label{fig:SuscCrystalsManyFields}
\end{figure}

We show the field-dependence of magnetization vs.\ temperature for Crystals \#1 and \#2 in Fig.\ \ref{fig:SuscCrystalsManyFields}. 
At higher field, the magnetic anomalies are diminished, while the N\'{e}el peak at 14.1 K remains present. In Fig.\ \ref{fig:SuscCrystalsManyFields}(a), data for crystal \#1 is shown for magnetic fields ranging from $\mu_0 H=0.25$ to 16 mT, and, strikingly, these curves overlap, showing linear-in-field behavior at low magnetic field. 
For field applied in-plane (Fig.\ \ref{fig:SuscCrystalsManyFields}(c)), the behavior is more complex, with the shoulder near 16 K no longer present by $\mu_0 H=5$ mT. On increasing from 5 to 50 mT, the magnetization at low temperature jumps to a new value, a sign of the in-plane spin flop transition reported to occur around 10 to 20 mT \cite{mcguireMagneticBehaviorSpinlattice2017,kuhlowMagneticOrderingCrCl31982}. 
For crystal \#2b under increasing out-of-plane field (Fig.\ \ref{fig:SuscCrystalsManyFields}(b)), there is also a shrinking of the 16 K shoulder, more prominent in the FC data (Fig.\ \ref{fig:SuscCrystalsManyFieldsCooling} in the Supplementary Materials \cite{supplement}.) 
For in-plane field, though, the magnetic anomalies in Crystal \#2 are dramatic, with a large peak at 16 K that diminishes rapidly on increasing magnetic field from $\mu_0 H=0.05$ to 2.0 mT.

\begin{figure}[t]
\begin{center}
\includegraphics[width=8.6cm]
{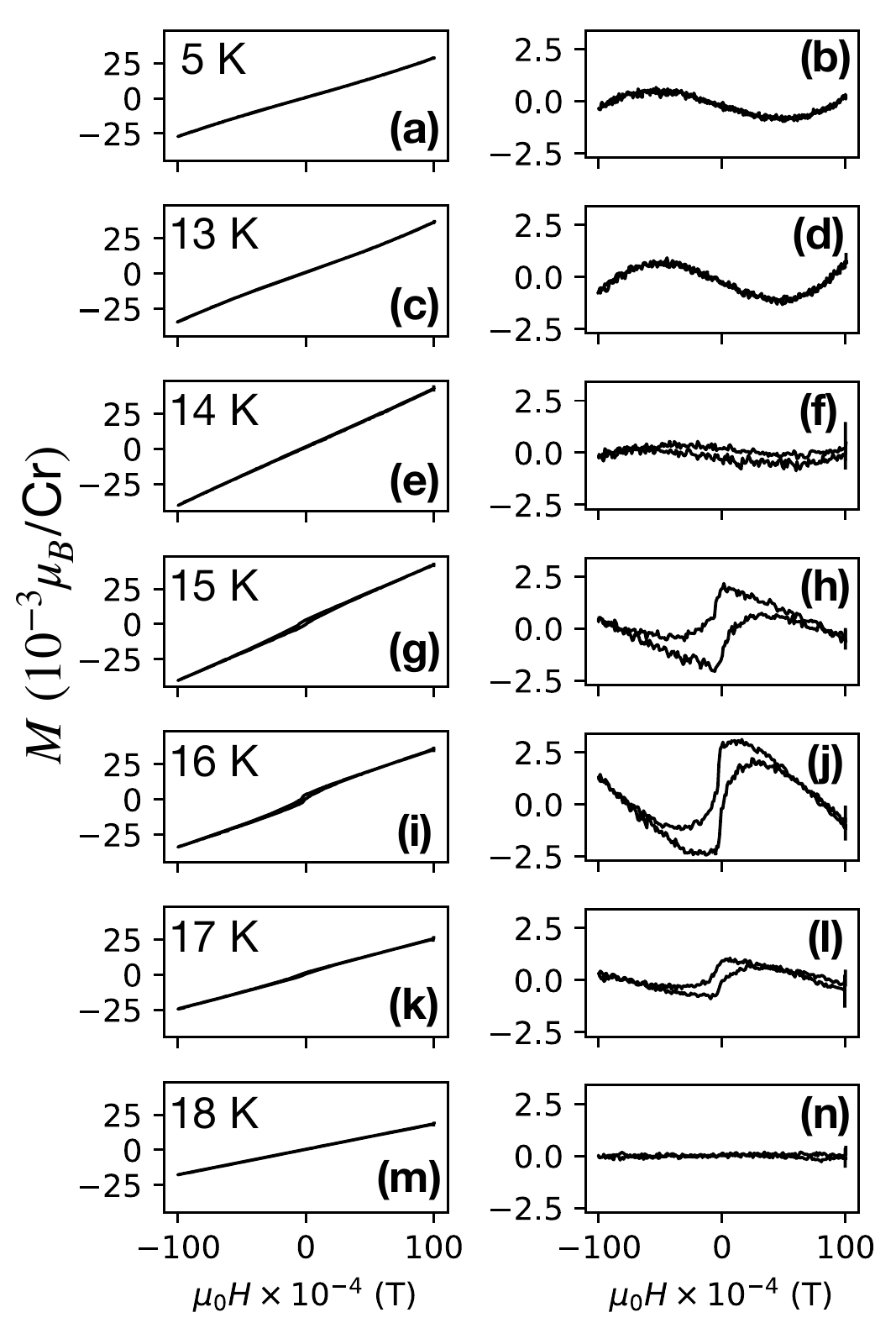}
\end{center}
\caption{(a,c,e,g,i,k,m) Isothermal magnetization vs.\ field hysteresis loops for Crystal \#2, with field applied in-plane. (b,d,f,h,j,l,n) The data from the first column, but with a line fitted and subtracted from the data for clarity.}
\label{fig:CrCl3MH}
\end{figure}

For a better look at the onset of the anomalous magnetic behavior, in Fig.\ \ref{fig:CrCl3MH} we show isothermal hysteresis loops of magnetization vs.\ field in the range $-0.01 \leq \mu_0 H \leq 0.01$ T. The first column shows the data. The second column shows the same data, except with a line fitted and subtracted to show the hysteresis behavior more clearly. A small hysteresis is present from about 14 to 17 K, with a coercive field of about 0.5 mT at 15 K. 
(An isothermal hysteresis at 14.8 K has been previously reported \cite{dehaasFurtherMeasurementsMagnetic1940}, though the hysteresis observed in that study was much wider (with a coercive field of $\sim$5 mT) than that seen in Fig.\ \ref{fig:CrCl3MH}(e) or in any other of our samples, including a pressed pellet.)
The curvature in the 5 and 13 K data (Fig.\ \ref{fig:CrCl3MH}(b,d)) is another indication of the in-plane anisotropy leading to the spin-flop transition, as can be seen more clearly for Crystal \#1 (see Supplemental Materials \cite{supplement}.) No hysteresis was seen for Crystal \#1. 

\subsection{Neutron scattering}
\begin{figure}[h]
\begin{center}
\includegraphics[width=8.6cm]
{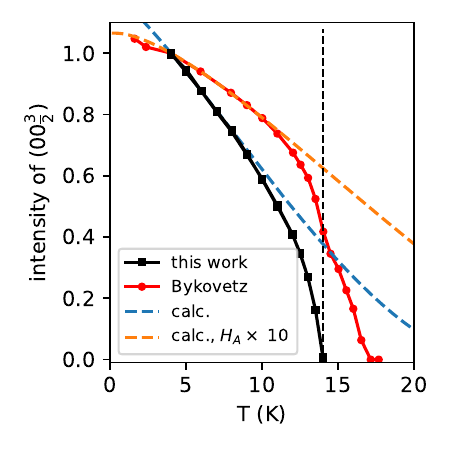}
\end{center}
\caption{The temperature dependence of the (squared) sublattice magnetization in terms of the neutron scattering intensity of the $(00\frac{3}{2})_R$ magnetic peak. Our data includes the integrated intensity of longitudinal scans along $(00\frac{3}{2})_R$, taken on a single crystal with minimal diffuse scattering or a monoclinic secondary phase. Data taken on both heating and cooling are shown, but overlap. Error bars are smaller than the markers. Data traced from Bykovetz, \emph{et al.}\ \cite{bykovetzCriticalRegionPhase2019} are also shown for comparison (without errorbars); only the data taken on warming are shown here, though they also reported cooling data, and a hysteresis was observed within 16 to 17 K. Two calculations of the sublattice magnetization squared are plotted as dashed lines, with the effective interlayer exchange $\mu_0 H_A$ set to 0.1 T or 1.0 T. All curves are normalized to their values at 4 K. The vertical dashed line indicates 14 K.
}
\label{fig:SublatticeMagnetization}
\end{figure}

Although the effect of M-type stacking defects on bulk magnetization is rather subtle, here, we show an example where their influence on bulk properties is very clear: the sublattice magnetization, as determined from the neutron scattering intensity of the magnetic $(00\frac{3}{2})$ peak. In Fig.\ \ref{fig:SublatticeMagnetization}, we show data taken via elastic neutron scattering measurements (on VERITAS) on a CrCl$_3$ single crystal with no diffuse scattering at low temperature (black squares), as well as data from Bykovetz, \emph{et al.} taken on a powder sample (red circles) \cite{bykovetzCriticalRegionPhase2019}. Both data sets show the integrated intensity of the $(00\frac{3}{2})$ peak as a function of temperature, normalized to their 4 K values. There is a clear difference between the two data sets. First, the intensity in our single crystal sample decreases more quickly, vanishing at $T_N = 14$ K; in contrast, the Bykovetz sample intensity decreases more slowly, with a kink at 14 K, but an ultimate transition temperature of 17 K. (Bykovetz, \emph{et al.}\ also noted a hysteresis on warming vs.\ cooling between 15.5 and 17 K \cite{bykovetzCriticalRegionPhase2019}; only their warming data is shown in Fig.\ \ref{fig:SublatticeMagnetization}. We have observed no hysteresis in our data.) 

We can compute the expected temperature dependence of the intensity of $(00\frac{3}{2})$ using a model from a series of nuclear magnetic resonance (NMR) studies on CrCl$_3$ \cite{narathNuclearMagneticResonance1961,narathLowTemperatureSublatticeMagnetization1963,narathSpinWaveAnalysisSublattice1965}. 
(See the Supplementary Materials for details \cite{supplement}.)
The intensity of $(00\frac{3}{2})$ is proportional to the sublattice magnetization squared, where the sublattice magnetization is the magnetization of one of the four symmetry-equivalent Cr ion sublattices of the rhombohedral structure of CrCl$_3$ (two for each honeycomb lattice, and two pairs of honeycomb lattices to capture the alternating direction of the spins layer-by-layer.) 
Against our data, we show the calculated squared sublattice magnetization using the model presented in Ref.\ \cite{narathSpinWaveAnalysisSublattice1965}; this model represents the interlayer coupling as part of an effective exchange field $\mu_0 H_A$, which we set to 0.1 T as observed in Ref.\ \cite{serriEnhancementMagneticCoupling2020}. 
There is good agreement between our data and the model up to $\sim$8 K, as seen in Fig.\ \ref{fig:SublatticeMagnetization}. (The model accounts for changes due to renormalization of spin waves, but is ultimately only applicable at sufficiently low temperature, with deviations above $\sim$8 K on comparison to NMR data \cite{narathSpinWaveAnalysisSublattice1965}.) 

On the other hand, for the data of Bykovetz, \emph{et al}.\ \cite{bykovetzCriticalRegionPhase2019}, a ten times larger anisotropy field, $\mu_0 H_A = 1.0$ T, was needed to get a good agreement with calculations. 
This agreement suggests that the roughly tenfold increase in interlayer magnetic coupling reported for M-type relative to R-type stacking \cite{serriEnhancementMagneticCoupling2020,kleinEnhancementInterlayerExchange2019} can be seen in the temperature dependence of the sublattice magnetization. 
(The agreement for a tenfold increase in $\mu_0 H$ is probably partially a coincidence. It seems unlikely that the sample measured in Ref.\ \cite{bykovetzCriticalRegionPhase2019} consisted of 100\% M-type stacking, though $\sim$75\% seems plausible in light of similar values seen in a ground powder sample of isostructural CrI$_3$ \cite{schneelochAntiferromagneticferromagneticHomostructuresDirac2024}.) Furthermore, the data of Ref.\ \cite{bykovetzCriticalRegionPhase2019} has an apparent transition temperature of 17 K, suggesting that 17 K is roughly the expected N\'{e}el temperature for a system with purely M-type stacking.

\begin{figure}[t]
\begin{center}
\includegraphics[width=8.6cm]
{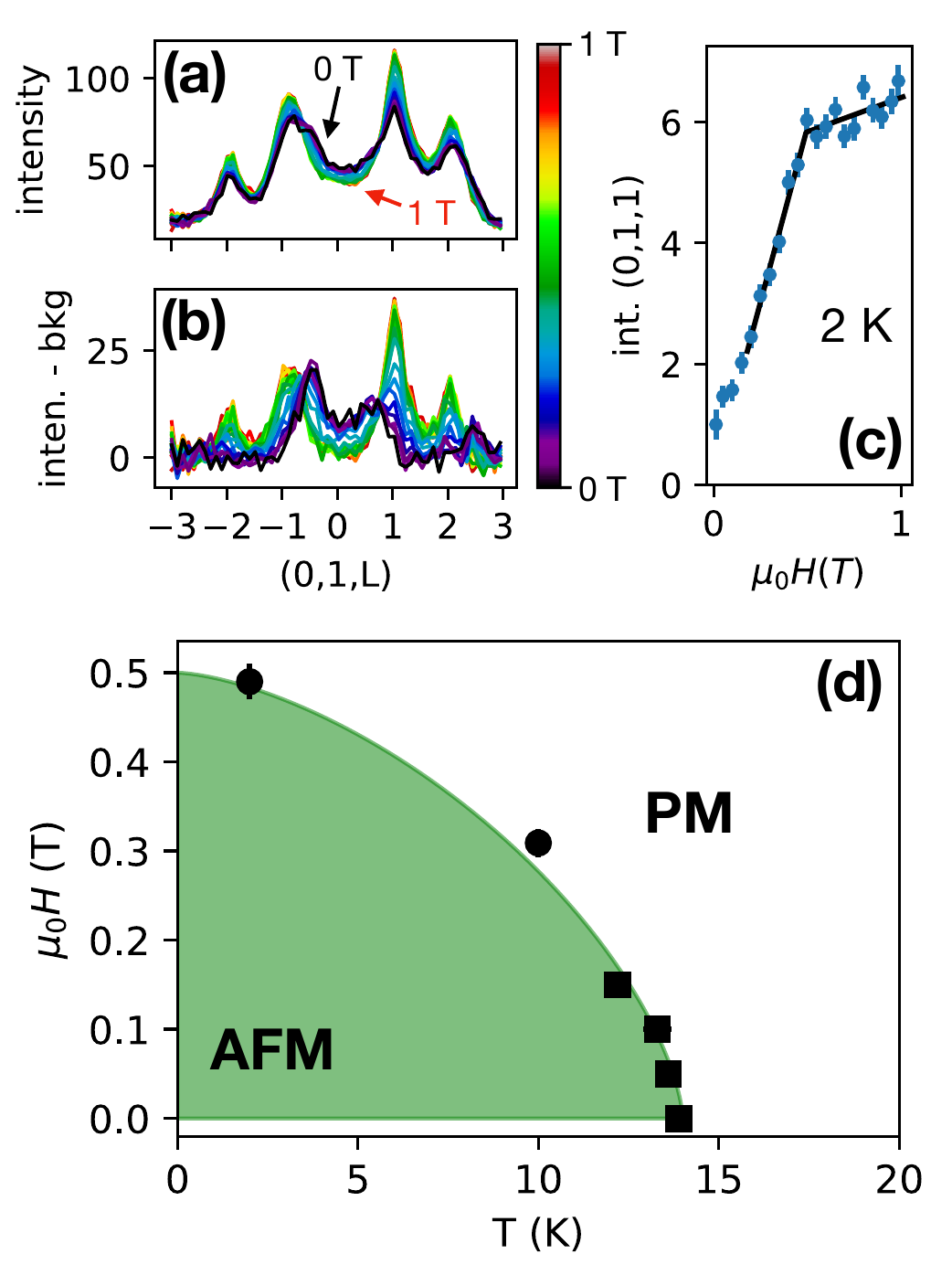}
\end{center}
\caption{(a) Data taken on a single crystal measured at CORELLI at 2 K, showing intensity along the $(0,1,L)$ line. Data for a range of applied magnetic fields are overplotted, ranging from $\mu_0 H=0$ T (black) to 1 T (red) as indicated by the colorbar. (b) The same data as in (a), except that a background of the intensity at 16 K was subtracted. (c) Intensity integrated within $0.9 \leq L \leq 1.1$ for each field at 2 K. A pair of connected line segments were fitted as shown to obtain a critical field of 0.49(2) T. (d) A phase diagram constructed from data taken with increasing field at constant temperature (circles), or constant field with increasing temperature (squares). Magnetic field not corrected for the demagnetization effect. Error bars in (d) are about the size of the markers.}
\label{fig:PhaseDiagram}
\end{figure}

A second crystal was used to construct a magnetic field-temperature phase diagram based on neutron scattering data taken at CORELLI. While several magnetic field-temperature phase diagrams have been reported in the literature, most were taken on few-layer flakes, and the phase diagram based on the specific heat data in Ref.\ \cite{mcguireMagneticBehaviorSpinlattice2017} did not show values below 10 K. 
Fig.\ \ref{fig:PhaseDiagram}(a) shows intensity along the $(0,1,L)$ line, with a background taken at 16 K subtracted in Fig.\ \ref{fig:PhaseDiagram}(b). For the R phase, nuclear Bragg peaks are expected at $L=\pm1$ and $\mp2$, with the signs corresponding to the two twins. The magnetic peaks in the AFM phase are expected at $L=\mp0.5$ and $L=\pm2.5$. For \ref{fig:PhaseDiagram}(a,b), data at 2 K are shown, taken at a series of magnetic fields applied along the $c$-axis from $\mu_0 H=0$ to 1 T. (To be precise, the magnetic field was ramped up continuously, and the resulting data were binned with respect to field.) The effect of an out-of-plane field is to rotate the spins out-of-plane, decreasing the AFM intensity at half-integer $L$ and increasing the FM component at integer $L$, resulting in the intensity changes seen in Fig.\ \ref{fig:PhaseDiagram}(b). We can track the transition by integrating intensity near half- or whole-integer $L$. For example, for the 2 K data in Fig.\ \ref{fig:PhaseDiagram}(c), intensity integrated within $0.9 \leq L \leq 1.1$ is shown as a function of field, increasing to about 0.5 T before exhibiting a discontinuous change in slope; a fit to a pair of connected line segments resulted in a critical field of $\mu_0 H_c = 0.49(2)$ T (consistent with data reported in Ref.\ \cite{cableNeutronDiffractionInvestigation1961}.) A similar analysis was done at 10 K. Values for $T_N$ at various applied fields (i.e., with constant field while temperature was continuously ramped up) were obtained for 0, 0.05, 0.1, and 0.15 T. The resulting phase diagram is shown in Fig.\ \ref{fig:PhaseDiagram}(d).

\section{Discussion}
\begin{figure}[h]
\begin{center}
\includegraphics[width=8.6cm]
{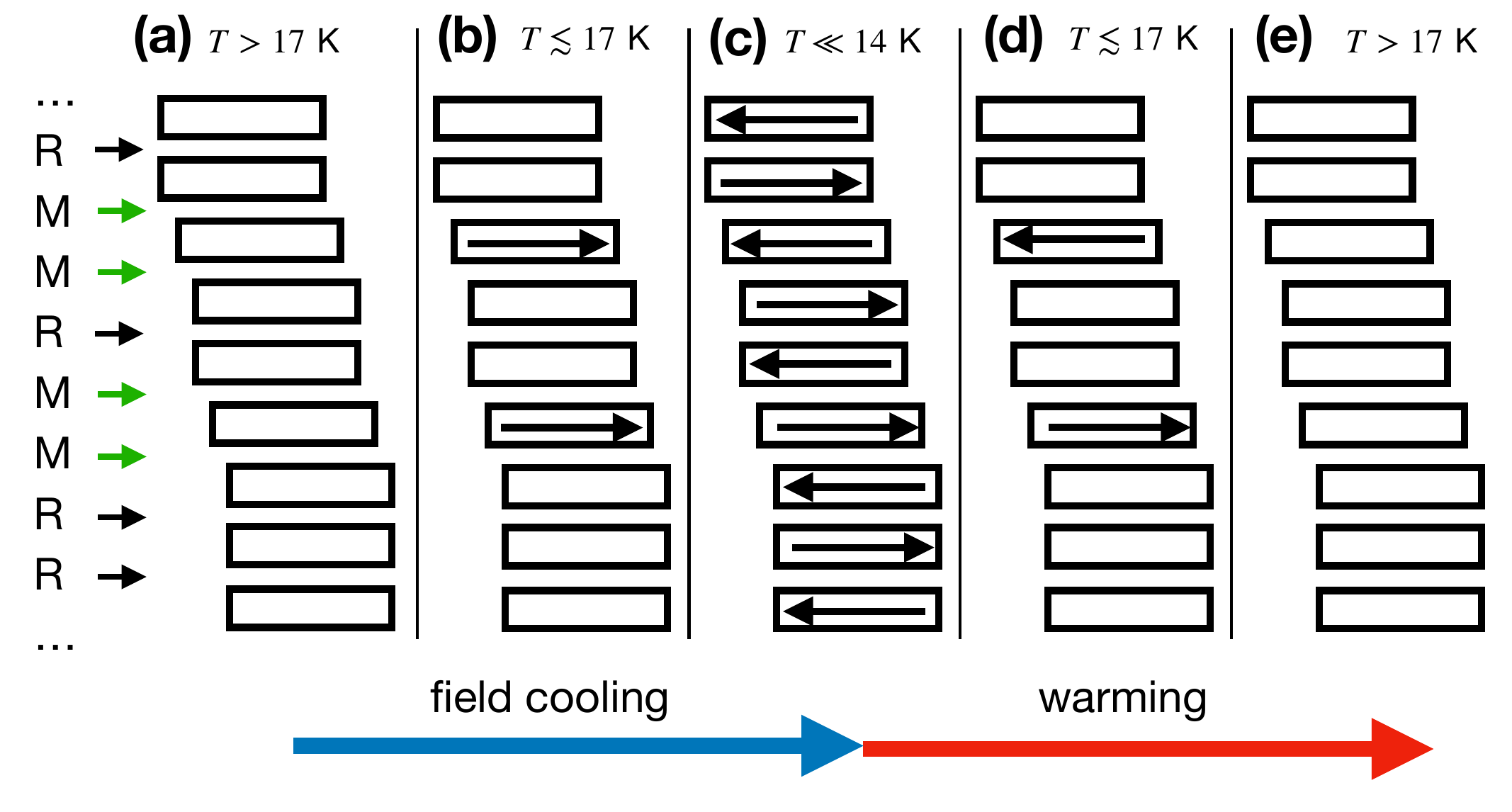}
\end{center}
\caption{An illustration of how the magnetization of individual domains may change on cooling and warming through 14 to 17 K, with a small (e.g., 0.25 mT) magnetic field applied throughout. (a) Initially, the sample is above 17 K when field is applied. (b) Below $\sim$17 K, layers in between M-type stacking boundaries, where the interlayer magnetic coupling would be about ten times stronger than for R-type stacking, become ordered. These isolated layers would tend to point in the same direction due to the field-cooling. Further cooling would expand these domains, but we assume that these domains would continue to be magnetically oriented along the direction of the field. (c) Below 14 K, full 3-dimensional order is achieved with alternating spin directions on each layer, and the orientation of the domains above 14 K is overwritten. (d) On subsequent warming above 14 K, the orientation of the few-layer domains near M-type stacking boundaries would match that of the low-temperature AFM order, pinned by intralayer magnetic anisotropy. (e) Above 17 K, order vanishes.}
\label{fig:FMDomains}
\end{figure}

We have shown that the low-field magnetic anomalies reported around 14 to 17 K \cite{liuAnisotropicMagnetocaloricEffect2020,bykovetzCriticalRegionPhase2019} are likely due to M-type stacking defects. These anomalies have hysteretic behavior, with a larger magnetization on field-cooling than ZFC warming. 
Below, we provide a rough explanation for this hysteretic behavior in terms of few-layer FM domains present a few K above $T_N$ and pinned by the in-plane anisotropy. Then, we will discuss the possible manifestation of the M-type stacking defects as the 17 K hump seen in in specific heat data, and ways in which our results may be informative to the study of other vdW-layered magnetic materials. 

We consider the situation depicted in Fig.\ \ref{fig:FMDomains}, where stacking disorder is preserved down to low temperature, increasing the local $T_N$ in regions with M-type stacking. 
If we apply a small field (e.g., 0.25 mT) and cool down to just below $\sim$17 K, we expect few-layer magnetic domains to form. 
A simple example is shown in Fig.\ \ref{fig:FMDomains}(b), where ordering occurs first within monolayers sandwiched by two M-type stacking boundaries. 
These ordered monolayers would constitute FM domains, and we would expect their orientation to correspond to the applied field.

On further cooling, we presume that these domains would spread out and continue to constitute FM (few-layer) domains oriented along the field. However, once these domains contact each other, the effect of the interlayer coupling should dominate and enforce 3-dimensional AFM order (Fig.\ \ref{fig:FMDomains}(c)), essentially ``erasing'' the previous orientations of the domains. 
This ``erasing'' can be seen in Fig.\ \ref{fig:SuscRepeated} in the Supplemental Material \cite{supplement}, in which magnetization is plotted for a sample over three heating/cooling cycles, showing that the hysteresis vanishes almost entirely below $\sim$14 K.

On warming above 14 K (Fig.\ \ref{fig:FMDomains}(d)), we presume that the few-layer FM domains would have their orientation pinned by the intralayer magnetic anisotropy, retaining the alignment of the 3-dimensional AFM order and, thus, having a much lower magnetization (ideally about zero) than on field cooling. Above $\sim$17 K, order disappears completely (\ref{fig:FMDomains}(e).)

This model may explain the hysteresis in the intensity of the magnetic $(00\frac{3}{2})$ Bragg peak reported in Ref.\ \cite{bykovetzCriticalRegionPhase2019}; on warming, although only certain layers may be magnetically ordered, their spin direction would still be compatible with that of the overall AFM order that existed below 14 K, in contrast to the situation on field-cooling where the alignment of the moments on different layers may be more incoherent due to being incompatible with the AFM order, resulting in a lower intensity.

The simple picture shown in Fig.\ \ref{fig:FMDomains} obscures complications which would require a more sophisticated model to explain. For instance, we can consider a monolayer FM domain that forms just below 17 K between two M-type stacking boundaries in an overall R-type region. On further cooling, the two neighboring layers may start to order, due to being in an environment of one M-type and one R-type stacking boundary (i.e., with an average interlayer coupling that is stronger than for purely R-type regions), and due to being next to an already-ordered monolayer domain. 
Thus, the initial monolayer FM domain would spread out and become a trilayer FM domain. The two new layers would, however, have moments in the opposite direction of the initial monolayer, but these moments would be less than that of the initial monolayer. Whether the net moment of the trilayer is still aligned along the field would depend on mathematical details and require a model that goes beyond the rough picture shown in Fig.\ \ref{fig:FMDomains}. 
More generally, the moments (and direction) of the few-layer FM domains, as well as the size of the magnetic anisotropy that may pin the moment direction of these domains, are quantities that would evolve with temperature, and their accurate accounting would require calculations that are beyond the scope of this paper. 

%

In light of M-type stacking defects in CrCl$_3$ explaining behaviors such as the low-field magnetic anomalies, the $\sim$2 T transition in isothermal magnetization-field measurements, and the temperature dependence of the $(00\frac{3}{2})$ magnetic Bragg peak, we consider the reported 17 K hump in specific heat data \cite{hansenHeatCapacitiesCrF31958,kostryukovaSpecificHeatAnhydrous1972,mcguireMagneticBehaviorSpinlattice2017}. 
The presence of the 17 K hump, in addition to the sharper 14 K peak seen in later studies \cite{mcguireMagneticBehaviorSpinlattice2017, bastienSpinglassStateReversed2019}, has been used as evidence for the existence of a ``ferromagneticlike'' phase \cite{mcguireMagneticBehaviorSpinlattice2017} in the CrCl$_3$ magnetic field-temperature phase diagram. 
On the one hand, there is an apparent inverse correlation between the size of the 17 K hump and 14 K peak (see Supplementary Materials \cite{supplement}), which suggests that the 17 K hump is due to the presence of M-type stacking defects. 
However, applied magnetic field appears to have opposite effects on these two features, moving the 14 K peak to lower temperature but apparently shifting the 17 K hump to higher temperature. 
We speculate that there are two distinct effects at play: 1) the ``background'' (i.e., the part of specific heat vs.\ temperature not including the 14 K peak or 17 K hump) shifts to higher temperature with field as the field modifies the spin correlations, and 2) if M-type stacking defects are present, a hump near 17 K is present, and expected to be suppressed (to lower temperature) with a field on the order of ten times that needed to suppress the 14 K peak. If the background does change substantially with field, then it would be hard to distinguish the simultaneous movement of the 17 K hump and the background in the data of Ref.\ \cite{mcguireMagneticBehaviorSpinlattice2017}.

We note that CrSBr is another vdW-layered compound with the same type of AFM order as CrCl$_3$ (spins aligned within each layer, antialigned between layers), and that an ``interlayer FM ordering'' phase has been claimed based on the same argument as in CrCl$_3$, in which different specific heat features move in different directions in response to applied field  \cite{leeMagneticOrderSymmetry2021}. 
Unlike CrCl$_3$, we are not aware of reported stacking defects in CrSBr, and its crystal structure does not appear to lend itself to such stacking variation. (For CrX$_3$ (X=Cl,Br,I), both M-type or R-type stacking involves the triangular anion lattices on the outside of each layer sitting on the corresponding anion lattices of neighboring layers in a similar way, even if the Cr$^{3+}$ honeycomb lattices are arranged differently. For the rectangular Br$^-$ lattice of CrSBr, though, no such variation is evident.)
A clear ``hump'' is not seen in the CrSBr specific heat data, but a gradual change in background with field can be seen \cite{leeMagneticOrderSymmetry2021}, with specific heat intensity moving toward higher temperature. 
Thus, we speculate that, in both CrSBr and CrCl$_3$, a field-induced shift in the background of specific heat data is an instrinsic behavior, and, in CrCl$_3$, the simultaneous shrinking of the 17 K hump and the movement of the background with increasing field may cause the appearance of the 17 K hump shifting to higher temperature with field even if the observed behavior is due to the background. 
It is not clear whether such intrinsic behavior is best explained by invoking an intermediate FM phase \cite{kuhlowMagneticOrderingCrCl31982,mcguireMagneticBehaviorSpinlattice2017,leeMagneticOrderSymmetry2021}, but, regardless, the potential for confusion from the effect of M-type stacking defects on specific heat data should not be overlooked.

Our findings on the effect of stacking defects on the properties of CrCl$_3$ may be useful in interpreting unusual behavior from other materials. 
For instance, FeBr$_2$ is another vdW-layered magnetic material (with AFM order similar to CrCl$_3$, except that the spins point perpendicular to the layers) where hysteretic low-field anomalies, present just below $T_N$, have been reported to be enhanced after grinding \cite{itoAnomalousMagneticBehavior1999}. Stacking defects have been implicated, likely CdCl$_2$-type within the overall CdI$_2$-type structure of FeBr$_2$ given the frequent presence of the CdCl$_2$- and CdI$_2$-type structures among the transition metal dihalides \cite{mcguireCrystalMagneticStructures2017}. It seems likely that weaker interlayer AFM coupling (``weaker'' since the anomalies are below $T_N$) would be found in bilayers of FeBr$_2$ with CdCl$_2$-type stacking if the same kind of few-layer measurements were done on FeBr$_2$ as on CrCl$_3$, but in the absence of such few-layer experiments, low-field magnetization measurements may prove insightful. 
Given that stacking defects are common in certain types of vdW-layered magnetic materials, it is important to understand the multitude of effects that stacking defects can cause. 

Both CrCl$_3$ and CrI$_3$ \cite{schneelochAntiferromagneticferromagneticHomostructuresDirac2024} are examples of systems where nanoscale behavior (the existence of different types of magnetic coupling across various stacking boundaries) can be seen in bulk properties such as magnetization measurements and neutron scattering. 
While, for both of these systems, nanoscale measurements indicated the presence of distinct kinds of interlayer magnetic coupling, our results show that a careful look at the bulk properties of materials with defects can provide similar insights into nanoscale interactions.

\section{Conclusion}
We have conducted low-magnetic-field susceptibility measurements on various CrCl$_3$ samples, showing that the anomalous behavior reported around 14 to 17 K is sample dependent and hysteretic. Samples with this anomalous behavior also show a transition around $\sim$2 T in isothermal magnetization-field measurements, as expected if M-type stacking defects were present given their reported tenfold greater interlayer magnetic coupling. Ground powder samples show especially strong anomalous behavior. 
We presented an explanation for the anomalous behavior just above 14 K in terms of few-layer FM domains. 
We argue that the effect of M-type stacking defects can be observed in the sublattice magnetization (via the neutron scattering intensity of the $(00\frac{3}{2})$ peak) and may also be responsible for variations in features observed in heat capacity measurements.

\section*{Acknowledgements} 

The work at the University of Virginia is supported by the Department of Energy, Grant number DE-FG02-01ER45927. A portion of this research used resources at the High Flux Isotope Reactor and the Spallation Neutron Source, DOE Office of Science User Facilities operated by Oak Ridge National Laboratory.

\nocite{narathNuclearMagneticResonance1961,shiraneNeutronScatteringTripleAxis2002}


%

\clearpage

\section*{Supplemental Materials}
\beginsupplement

\subsection{Comparison of specific heat data from the literature}

\begin{figure}[h!]
\begin{center}
\includegraphics[width=8.6cm]
{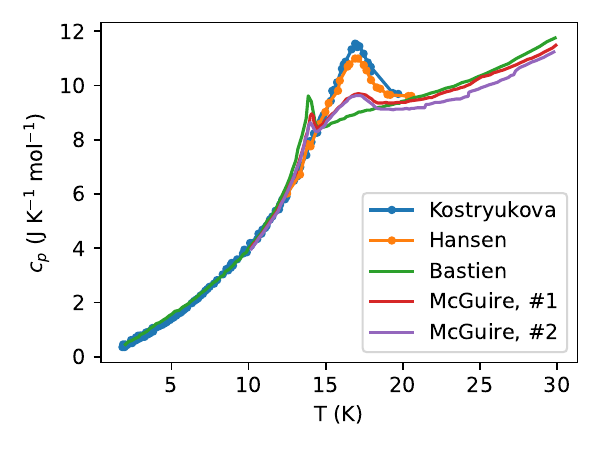}
\end{center}
\caption{Specific heat data traced from various publications labeled by first author (Kostryukova and Luk'yanova   \cite{kostryukovaSpecificHeatAnhydrous1972}, 
Hansen and Griffel \cite{hansenHeatCapacitiesCrF31958}, 
Bastien, \emph{et al.}\ \cite{bastienSpinglassStateReversed2019}, and McGuire, \emph{et al.}\ (two samples) \cite{mcguireMagneticBehaviorSpinlattice2017}.) }
\label{fig:CpComparison}
\end{figure}

Here, we show specific heat data presented in the literature, labeled in Fig.\ \ref{fig:CpComparison} by the first authors of Refs.\ \cite{kostryukovaSpecificHeatAnhydrous1972,hansenHeatCapacitiesCrF31958,bastienSpinglassStateReversed2019,mcguireMagneticBehaviorSpinlattice2017}. The data of Kostryukova, \emph{et al.}\  \cite{kostryukovaSpecificHeatAnhydrous1972} and Hansen, \emph{et al.}\  \cite{hansenHeatCapacitiesCrF31958} were taken on pressed pellet samples, which, presumably, would have inhibited $M$$\rightarrow$$R$ transitions on cooling. This probably explains the absence in the data of Kostryukova, \emph{et al.}\  \cite{kostryukovaSpecificHeatAnhydrous1972} of the sharp peak near 14 K first observed by McGuire, \emph{et al.}\ \cite{mcguireMagneticBehaviorSpinlattice2017}. Notably, the hump near 16-17 K is much larger in the data of Kostryukova, \emph{et al.}\ \cite{kostryukovaSpecificHeatAnhydrous1972} and Hansen, \emph{et al.}\  \cite{hansenHeatCapacitiesCrF31958}.
The data of Bastien, \emph{et al.}\  \cite{bastienSpinglassStateReversed2019} show an even larger 14 K peak and an even more diminished 16-17 K hump. Overall, these data suggest an inverse correlation between the size of the 14 K peak and the 17 K hump, which suggests that the 17 K hump is due to a sample-dependent factor such as the presence of M-type stacking defects.

\clearpage

\subsection{Additional magnetization data}
\begin{figure}[h!]
\begin{center}
\includegraphics[width=8.6cm]
{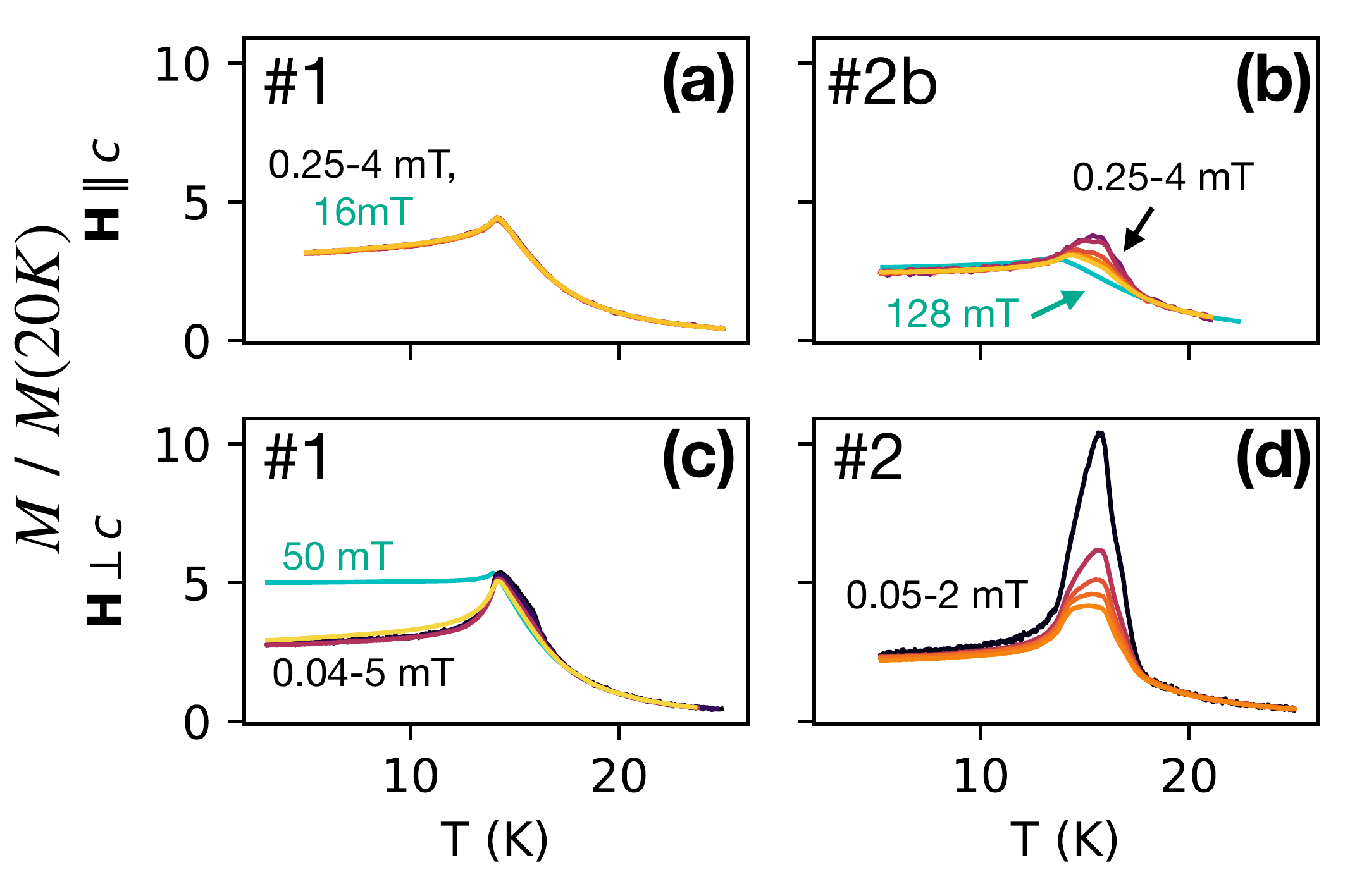}
\end{center}
\caption{Magnetization vs.\ temperature at many applied magnetic fields. 
Fields from $\mu_0 H=0.04$ to 5 mT are shown on a color scale from black to yellow. The plotted range of these fields is $\mu_0 H=0.25$ to 4 mT for (a), 0.25 to 4 mT for (b), 0.04 to 5 mT for (c), and 0.05 to 2 mT for (d). 
Additionally, the cyan curves show data taken at a higher magnetic field, specifically, 16 mT for (a) (overlapped by the other data), 128 mT for (b), and 50 mT for (c). All data are FC and normalized to their value at 20 K.}
\label{fig:SuscCrystalsManyFieldsCooling}
\end{figure}

Figure \ref{fig:SuscCrystalsManyFieldsCooling} shows the FC magnetic susceptibility data corresponding to the ZFC data shown in Fig.\ \ref{fig:SuscCrystalsManyFields}. 

\begin{figure}[h]
\begin{center}
\includegraphics[width=8.6cm]
{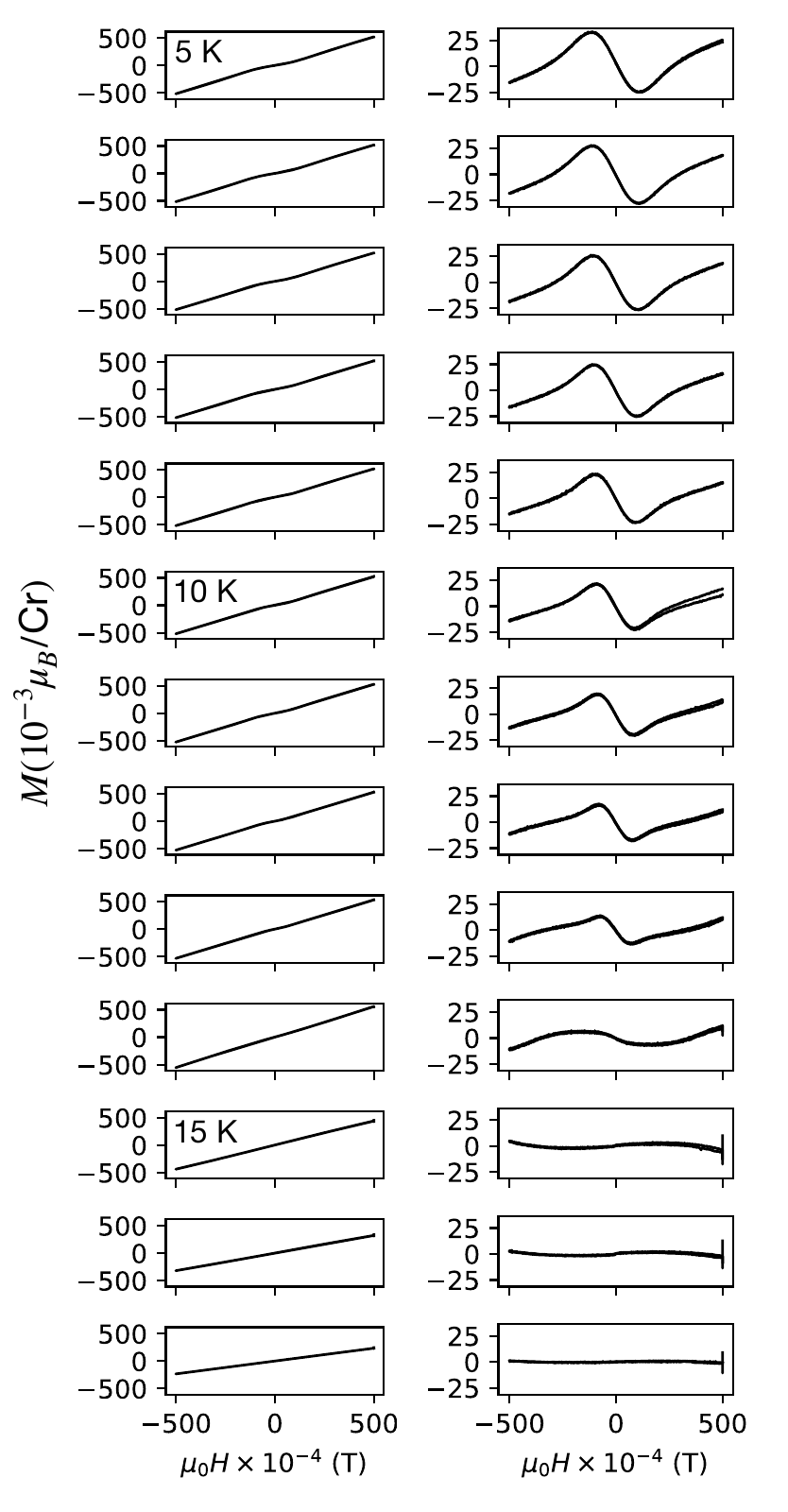}
\end{center}
\caption{Magnetization vs.\ field hysteresis loops for Crystal \#1 in the range $-0.05 \leq \mu_0 H \leq 0.05$ T. Each row corresponds to a different temperature, decreasing in 1 K steps from 5 K to 17 K. The first column shows the raw data, and the second column shows the data with a line fitted and subtracted.}
\label{fig:CrCl3MHSup}
\end{figure}

In Fig.\ \ref{fig:CrCl3MHSup}, we show isothermal magnetization-field hysteresis loops for Crystal \#1 as a counterpart to the Crystal \#2 data shown in Fig.\ \ref{fig:CrCl3MH}.

\begin{figure}[h]
\begin{center}
\includegraphics[width=8.6cm]
{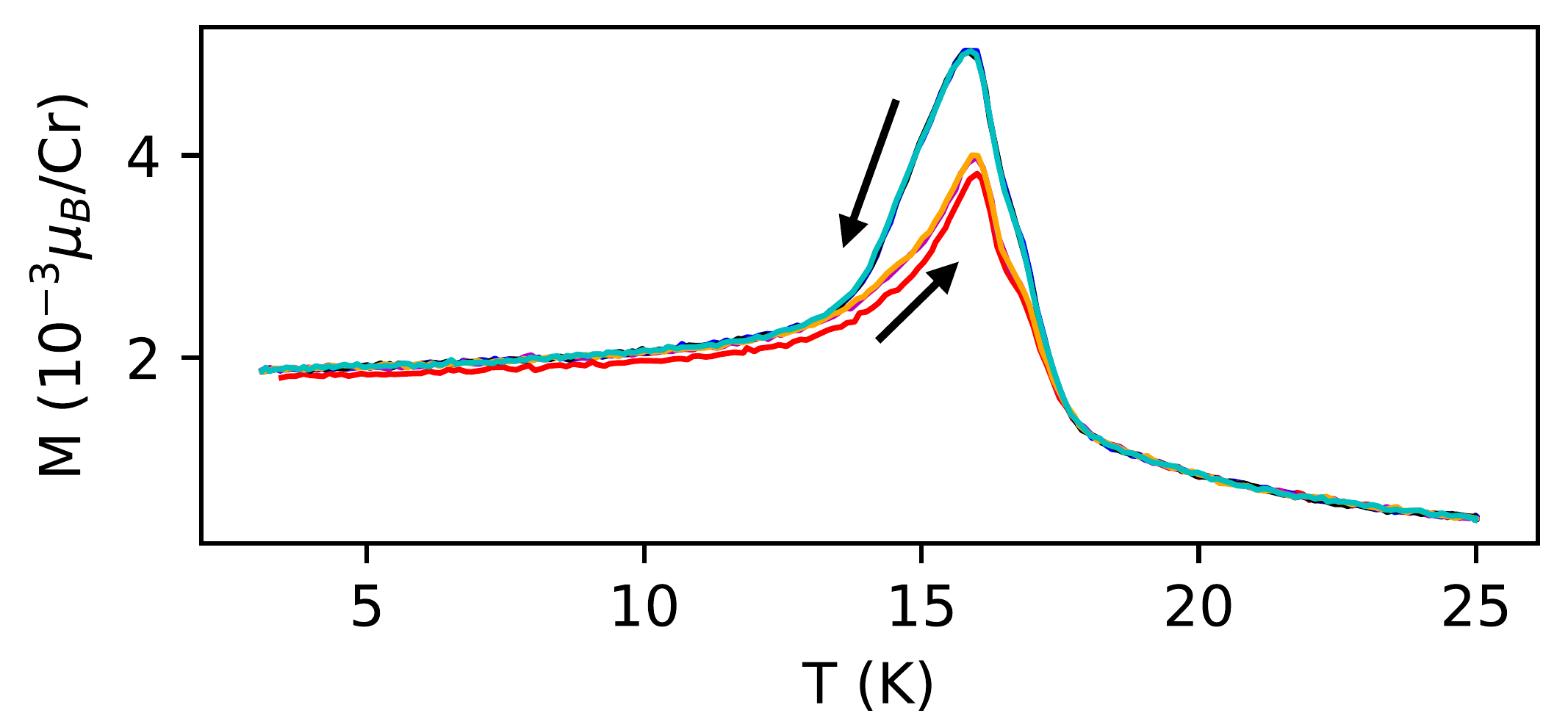}
\end{center}
\caption{Magnetization vs.\ temperature for one of the ground powder samples over three heating/cooling cycles. The nominal field was $\mu_0 H = 2.5$ mT. The red curve is the initial (ZFC) magnetization, taken on warming. The first cooling curve is dark blue, but is overlapped by the other two cooling curves (black and cyan.) The 2nd and 3rd heating curves (orange and magenta) also overlap.}
\label{fig:SuscRepeated}
\end{figure}

Fig.\ \ref{fig:SuscRepeated} shows data taken over three heating/cooling cycles for a ground powder sample. The red curve shows the ZFC (warming) curve, followed by cooling and warming curves which overlap. We see that there is minimal difference between warming data taken after applying field for the first time or after having cooled in field, suggesting that once the sample is below $\sim$14 K and 3-dimensional AFM order sets in, thermal history typically becomes irrelevant.

\clearpage

\subsection{Comparison between low-field $M$ vs.\ $T$ data and low-temperature $M$ vs.\ $H$ data}

\begin{figure}[h]
\begin{center}
\includegraphics[width=8.6cm]
{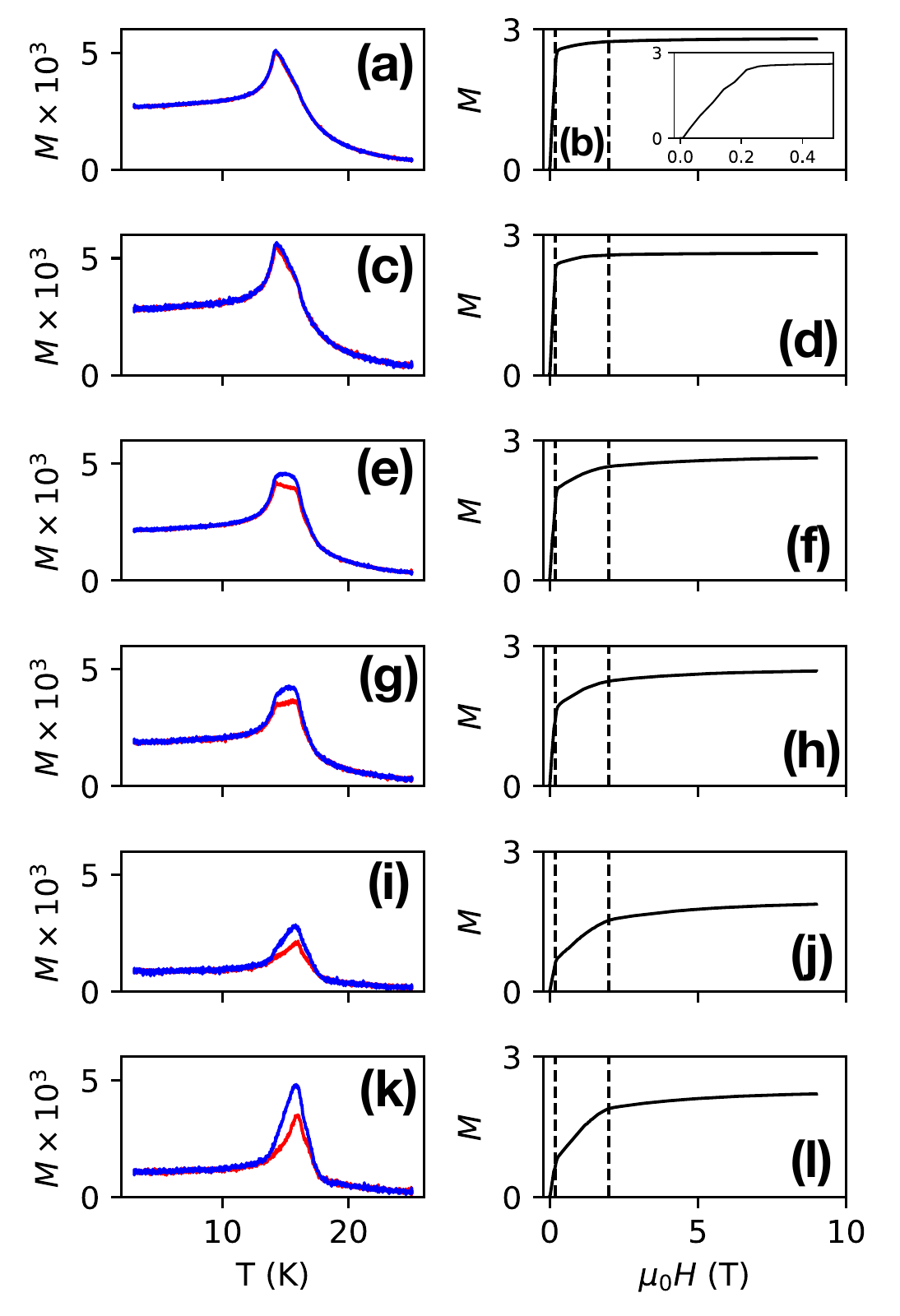}
\end{center}
\caption{Magnetization data for six crystals, with field applied in-plane. The first column shows low-field (nominally $\mu_0 H=0.25$ mT) magnetization vs.\ temperature data, with ZFC (blue) measurements followed by FC (red) data. The second column shows magnetization vs.\ applied field data taken at 3 K for the corresponding sample. The inset of (b) shows a zoomed-in version of that plot. Magnetization is in units of $\mu_B$/Cr. Vertical lines denote $\mu_0 H=$ 0.2 and 2 T. The data in (a,b) and (k,l) are also shown in the main text in Fig.\ \ref{fig:grinding}(b-e).}
\label{fig:ManyMH1}
\end{figure}

\begin{figure}[h]
\begin{center}
\includegraphics[width=8.6cm]
{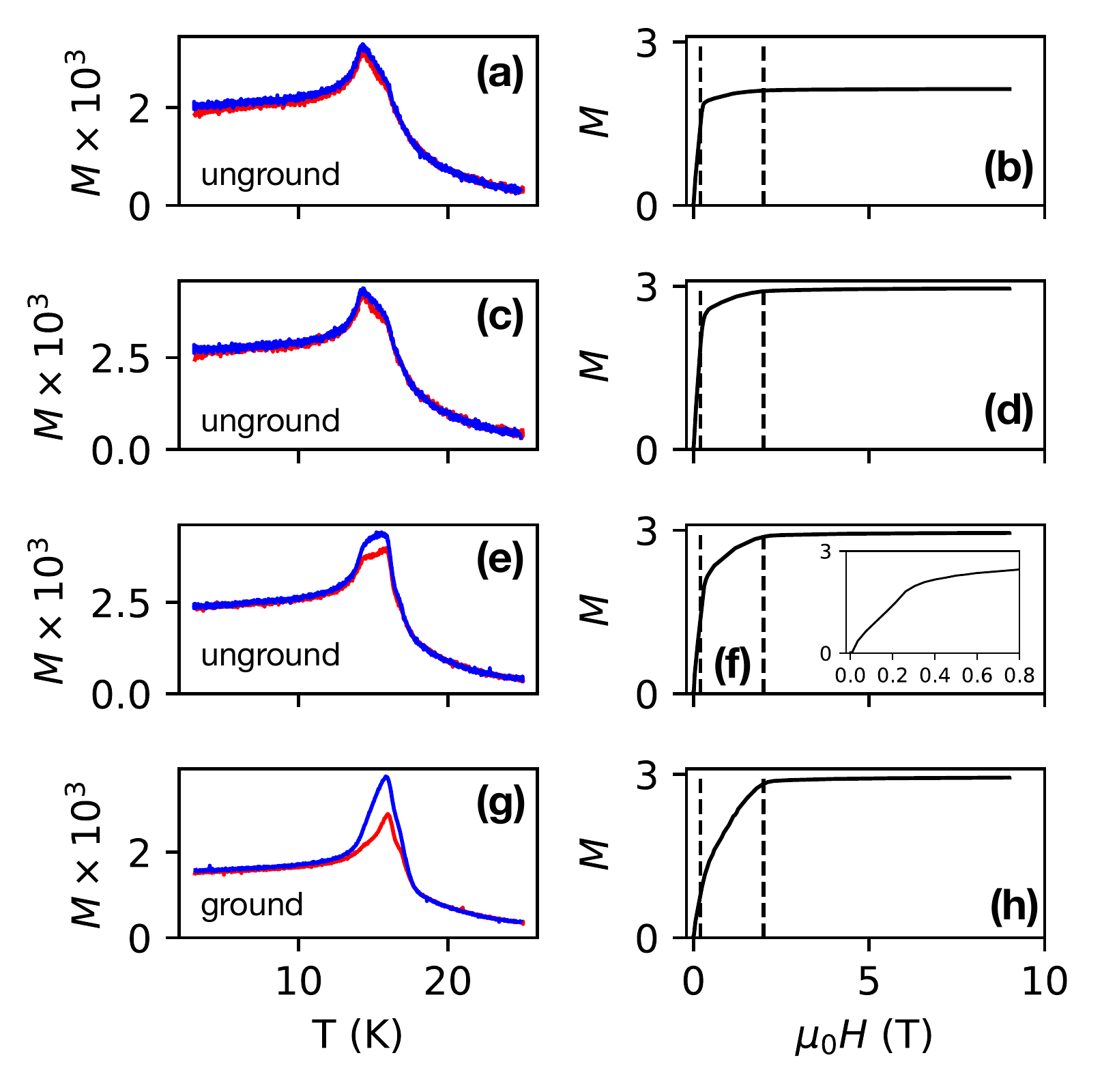}
\end{center}
\caption{Magnetization data for four powder samples. The first column shows low-field (nominally $\mu_0 H=0.25$ mT) magnetization vs.\ temperature data, with ZFC (blue) measurements followed by FC (red) data. The second column shows magnetization vs.\ applied field data taken at 3 K for the corresponding sample. The inset of (f) shows a magnified version of that plot. Magnetization is in units of $\mu_B$/Cr. Vertical lines denote $\mu_0 H=$ 0.2 and 2 T.}
\label{fig:ManyMH2}
\end{figure}

In Fig.\ \ref{fig:ManyMH1}, we show magnetization data for six crystals, with field applied in-plane. The first column shows low-field temperature dependence, with a hysteresis within 14 to 17 K present for the bottom four rows. The second column shows magnetization vs.\ field taken at 3 K for the corresponding samples in the first column. All crystals show a transition at $\mu_0 H \approx$ 0.2 T, as expected for the R phase \cite{mcguireMagneticBehaviorSpinlattice2017,serriEnhancementMagneticCoupling2020}. The bottom four crystals also show a transition near 2 T, as reported for M-type stacking \cite{serriEnhancementMagneticCoupling2020}. Overall, we see that the 2 T transition becomes more prominent the larger the low-field hysteresis anomaly becomes. 

Fig.\ \ref{fig:ManyMH2} shows similar data as for Fig.\ \ref{fig:ManyMH1}, except for powder samples rather than single crystals. Due to the demangetization effect, the field needed to suppress the AFM order (assuming a pure R-phase crystal) can be as high as $\sim$0.6 T \cite{mcguireMagneticBehaviorSpinlattice2017} when field is applied out-of-plane. In practice, we see the transitions occur at only slightly higher fields than the $\sim$0.2 and 2 T values seen for single crystals with field applied in-plane. 
Regardless, the trend is similar as for the single-crystal data in Fig.\ \ref{fig:ManyMH1}, where, e.g., the sample with the most prominent hysteresis (Fig.\ \ref{fig:ManyMH2}(g)) has the most prominent 2 T transition (Fig.\ \ref{fig:ManyMH2}(h)). 

\clearpage

\subsection{Temperature dependence of the sublattice magnetization}

In Fig.\ \ref{fig:SublatticeMagnetization} we showed the integrated intensity of the magnetic peak at $(00\frac{3}{2})$ as measured by longitudinal neutron scattering scans on a single crystal. The intensity of $(00\frac{3}{2})$ is proportional to the sublattice magnetization squared. 
In detail, the AFM structure of CrCl$_3$ consists of four sublattices, with two sublattices per honeycomb lattice in each layer, and two sets of honeycomb lattices on alternating layers to account for the alternating orientation of the spins \cite{narathSpinWaveAnalysisSublattice1965}. 
The sublattice magnetization thus represents the average magnetization within a layer.
The neutron scattering intensity of a magnetic peak at $(HKL)$ is given by $\sqrt{\vec{F}_M \cdot \vec{F}_M^*}$, where $\vec{F}_M$ is the magnetic structure factor, given by \cite{shiraneNeutronScatteringTripleAxis2002}
\begin{equation}
\vec{F}_M = \sum_j p_j \vec{S}_{\perp j} e^{i 2 \pi (H x_j + K y_j + L z_j)}. 
\end{equation}
We have neglected the Debye-Waller factor, which would show minimal change in the temperature range of interest. The quantity $p_j$ includes the magnetic form factor.  The sum runs over the magnetic unit cell, which we define using the same coordinates $(x_j,y_j,z_j)$ as for the structural unit cell but with a doubled range in the out-of-plane direction (i.e., $0 \leq z_j < 2$) to account for the antiferromagnetism. 
$\vec{S}_{\perp j}$ denotes the component of $\vec{S}_j$ that is perpendicular to the momentum transfer $\vec{Q}$; for $(00\frac{3}{2})$, $\vec{S}_{\perp j} = \vec{S}_j$. 
The intensity of $(00\frac{3}{2})$ is simply proportional to $|\vec{S}_{\perp j}|^2 = S^2$, and thus the sublattice magnetization squared.

We compared our data with calculations of the sublattice magnetization from the model (Eq.\ 5.8) in Ref.\ \cite{narathSpinWaveAnalysisSublattice1965}. 
This model accounts for the decrease in the sublattice magnetization on heating in terms of the excitation of magnons. 
Due to the simplicity of the magnetic structure of CrCl$_3$, only three parameters were needed for quantitative agreement below $\sim$8 K with the temperature dependence of an NMR resonance frequency: the nearest-neighbor (intralayer) exchange coupling $J_T$, an effective anisotropy field $H_A$ that can account for the interlayer coupling, and a scale factor \cite{narathNuclearMagneticResonance1961,narathLowTemperatureSublatticeMagnetization1963,narathSpinWaveAnalysisSublattice1965}. 
The model also includes renormalized spin-wave energies \cite{narathSpinWaveAnalysisSublattice1965}.
(While Refs.\ \cite{narathNuclearMagneticResonance1961,narathLowTemperatureSublatticeMagnetization1963} entirely accounted for the interlayer coupling via $H_A$, the model of Ref.\ \cite{narathSpinWaveAnalysisSublattice1965} also included an interlayer exchange coupling $J_L$. However, we find that $J_L$ is not necessary for good agreement with our data, and the studies reporting the roughly tenfold increase in interlayer coupling for M-type stacking \cite{serriEnhancementMagneticCoupling2020, kleinEnhancementInterlayerExchange2019} both interpreted their observed increases in $H_A$ entirely as increases in the interlayer coupling, in the process assuming other sources of anisotropy to be negligible. For consistency with these studies, we set $J_L=0$ and assumed $H_A$ encapsulates the effects of interlayer magnetic coupling.)
We set $\mu_0 H_A = 0.1$ T (the value reported in Ref.\ \cite{serriEnhancementMagneticCoupling2020}. The nearest-neighbor (intralayer) exchange coupling $J_T$ was kept at 0.90 meV per bond. The resulting calculated curve agrees well with our data below $\sim$8 K. For comparison with the data of Ref.\ \cite{bykovetzCriticalRegionPhase2019}, a tenfold larger effective anisotropy field was used, $\mu_0 H_A = 1.0$ T.

\end{document}